\documentstyle{sup}
\input psfig
\vspace{-8pt}
\author%%%[Superlattices and Microstructures, Vol.\ xx, No.\ x, 199x]
[G{\"o}teborg Preprint APR 97-22]
{Mats~Jonson$^1$, Ilya~V.~Krive$^{1,2}$,
Peter Sandstr{\"o}m$^1$, and Robert~I.~Shekhter$^1$
\\
{\normalsize\it $^1$Department of Applied Physics,}\cr
{\normalsize\it Chalmers University of Technology
and G{\"o}teborg University, S-412 96 G{\"o}teborg, Sweden}\\
{\normalsize\it and}\\
{\normalsize\it $^2$B. Verkin Institute for Low Temperature Physics and 
Engineering,}\cr
{\normalsize\it Academy of Sciences of Ukraine, 310164 Kharkov, Ukraine}
}
\title%%%[Superlattices and Microstructures, Vol.\ xx, No.\ x, 199x]
[Nonlinear Conductance of Nanowires]
{Nonlinear Conductance of Nanowires --\\A Signature of
Luttinger Liquid Effects?}
\pagerange{\pageref{firstpage}--\pageref{lastpage}}

\def\LaTeX{L\kern-.25em\raise.425ex\hbox{a}\kern-.075em\TeX}

\def\fakebold#1{\relax\ifvmode\leavevmode\fi%
\ifmmode%
\setbox0=\hbox{$#1$}%
\else%
\setbox0=\hbox{#1}%
\fi%
\kern-.02em\copy0 \kern-\wd0%
\kern .04em\copy0 \kern-\wd0%
\kern-.0125em\raise.02em\box0%
}%
\begin{document}
\label{firstpage}
\maketitle
\sloppy
\begin{center}
\received{(%Received 
25 October 1997)}
\end{center}
%**********************************************************************%
%
\begin{abstract}
We analyze recent measurements of the room temperature 
current - voltage  characteristics
of gold nanowires, whose zero current conductance is quantized in units
of $2e^2/h$. A faster than linear increase of current with voltage was
observed at low voltages beginning from $V_c=0.1$~V. We analyze the
nonlinear behavior in terms of a dynamic Coulomb blockade of conducting
modes and show that it may be explained as a Luttinger-liquid effect.
\end{abstract}

\section{Introduction}

Recently conductance quantization at room temperature
was discovered \cite{CQM} in metallic nanowires formed by breaking a contact
between two metallic electrodes. Although the accuracy of the quantization
in metallic nanowires as a rule is less then what can be achieved in
ballistic nanocontacts formed in the two-dimensional (2D) electron gas of 
gated semiconductor heterostructures \cite{CQS}, the mere fact that
plateaus appear at quantized values of conductance $ G \simeq n(2e^2
/h), (n = 1,2,\ldots) $ shows that the electron transport through
the nanowire is ballistic (the mean free path of charge carriers
is larger than the length $L$ and width $W$ of the wire). In thin
metallic nanowires the screening of ionic potentials by
mobile charge carriers, which is very efficient
in the bulk electrodes, could be considerably suppressed. Hence, the
electrostatic potential felt by electrons moving along the wire becomes
more smooth than is the case in bulk metals. In this picture the 
conductance quantization at room
temperature can be understood within the framework of the
``conductance is transmission'' - picture --- originally proposed by Rolf 
Landauer \cite{Rolf} --- applied to a ballistic contact with almost perfect 
transmission. The approach of Landauer was subsequently developed by him and 
others into what we now know as the Landauer - B{\"u}ttiker formalism, 
which is what is used below (see e.g. the review in Ref.~\cite{Land_Butt}).

Suppression of screening in metallic nanowires implies in addition that
the effects of electron-electron interactions could be important
for the nature of electron transport through a nanocontact. It
is well known that for 1D ballistic wires with a length that is effectively
infinite, the interaction renormalizes the conductance quantum to a value which
depends on the interaction strength (in the Tomonaga-Luttinger liquid
model the conductance per spin orientation is $ K_{\rho}(e^2/h)$,
where $ K_{\rho} $ is the charge correlation parameter of the Luttinger
liquid \cite{Apel_Rice,KF1}). However, the conductance of a {\em finite}
ballistic wire attached to leads with noninteracting electrons
takes on integer values (in terms of the conductance quantum $e^2/h$) 
regardless
of the interactions in the wire \cite{Maslov_Stone}. Therefore, the effects 
of electron-electron interactions are not revealed in the linear response
regime for a finite ballistic (reflectionless) contact.

If the electron transport through a wire is not purely ballistic
(i.e. there is a finite probability for electrons to be
backscattered by impurity- or channel imperfections) the above
picture is radically changed. In an effectively infinite 1D wire
even arbitrarily small backscattering suffices to block the dc-current at 
vanishingly low voltages and temperatures
\cite{KF2,GRS}. The effect is due to a strong enhancement of the bare
backscattering amplitude by charge fluctuations at long distances.
For a finite wire the renormalized barrier is also finite; the
procedure of barrier renormalization should be stopped when a low-energy
scale of order $e^2/L$ is reached. 
Therefore, in a multichannel quantum wire
the transmittivity of quantized modes will depend both on their
bare transmission coefficients and on the strength of electron-electron
interaction. The transmitting modes most sensitive to backscattering
(the criteria will be given below) are converted to reflected
modes by interaction effects. In other words the renormalized
barrier is the actual barrier experienced by electrons propagating
along the wire and this very barrier separates transmitting- from
reflected quantized modes. The blockade imposed by the
electron-electron interaction is gradually lifted by the bias voltage
beginning from $\; V > V_c\simeq e/L\;$ resulting in a power-law
$I-V$ characteristics.

Nonlinear $I-V$ characteristica for metallic nanowires in
a regime where the linear conductance was quantized have recently been
measured by Costa-Kr{\"a}mer {\em et al.} \cite{Garcia}.
The following remarkable universal properties were observed:
\begin{enumerate}
\renewcommand{\theenumi}{(\roman{enumi})}
\item the characteristic crossover voltage $V_c \sim 0.1$~V to the nonlinear
regime is  sample-independent and anomalously small in comparison with the 
atomic energy scale $\sim$~1~eV set by the spacing of quantized mode energies
in the contact \cite{channels}
\item the nonlinear character of the $I-V$ curves does not depend on the
conductance in the zero voltage limit and is therefore independent 
of the number of conduction channels
\item  the nonlinear contribution to 
the current is approximately proportional to the third power of the voltage
\end{enumerate} 

The purpose of this work is to show that all the above listed
properties can be explained (at least qualitatively)
if one associates the nonlinear current
with the current produced by one
(or possibly a few) --- partly reflected --- 
modes in a multichannel wire of interacting electrons. The current associated
with this mode, which is small due to ``Coulomb
blockade" effects in the linear response regime, is strongly enhanced when the 
bias voltage is increased; eventually the voltage lifts the blockade
and converts the reflected mode to an almost perfectly transmitting one.
As we proceeed to elaborate this conjecture we recall first how the 
universal power-law dependence of current on voltage
arises in the Luttinger liquid approach. To be concrete we consider a 1D model
of interacting spinless electrons scattered by a local potential barrier.
%The model can be treated analytically in two opposite limits --
%weak and strong interaction.

\section{Luttinger liquid model for nanowires -- Weak interaction limit}
For our purpose it is important to be
able to describe a situation where the bare transmission 
probability can be close to --- but not necessarily equal to ---  unity so that
the conductance increases by essentially one quantum when the Luttinger 
liquid 
%(Coulomb blockade)
 effects are suppressed (for example by a high bias
voltage). This opening up of a new conduction channel is precisely what 
we propose may explain the nonlinearities in the measured $I-V$ curves.
For arbitrary barriers the model can be treated analytically only
for weakly interacting electrons when the effect of interaction can
be reduced to a barrier renormalization produced by the formation
of Friedel oscillations of the electron 
 density near the barrier.
The renormalized transmission 
coefficient is \cite{Matveev_Glazman_Yue}

\begin{equation}
T^R(\varepsilon) = \frac{T_0\left(\frac{\varepsilon -\varepsilon_F}{D_0}\right)
^{2\gamma}}{R_0 + T_0\left(\frac{\varepsilon -\varepsilon_F}{D_0}
\right)^{2\gamma}} .
\end{equation}
Here $T_0 (R_0)$ is the bare transmission (reflection) coefficient
($T_0 + R_0 = 1$), $\varepsilon_F$ is the Fermi energy, $D_0$ is an ultraviolet
cutoff (see below), $\;\gamma 
\equiv \gamma_f -\gamma_b = [V_{ee}(0) - V_{ee}(2k_F)]/2\pi\hbar
v_F \;$ is a dimensionless parameter which characterizes the strength
of the electron-electron interaction, and $ V_{ee}(q)$ is the Fourier 
transform of the interaction potential.
Strictly speaking the parameter $\gamma$ should be small
for this approach to be valid
($\gamma\simeq e^2/2\pi\hbar v_F \ll 1$), but it is believed that
Eq.~(1) still holds qualitatively in the range $\gamma \sim 1$. In what
follows we will regard $\gamma$ as an input parameter.

The renormalized transmission coefficient (1) leads to an expression for the 
current when put into Landauer's formula. At zero temperature one finds

\begin{equation}
I(V) = \frac{e}{h}\int_{\varepsilon_F}^{\varepsilon_F
+ eV} d\varepsilon \;T^R(\varepsilon - \varepsilon_F)
 \equiv \frac{e}{h}F_\gamma\left(\frac{eV}{D_0}
\left[\frac{T_o}{R_0}\right]^{\frac{1}{2\gamma}}\right) .
\label{two}
\end{equation}
We assume that the bare transmission amplitude, $T_0$, is
energy independent near the Fermi level, so that the entire energy
dependence of $ T^R(\varepsilon) $ stems from the universal effects
of barrier renormalization. The function $F_\gamma(x)$ has to be obtained numerically except in special cases (see below).

For a small renormalized transmission
probability ($T^R \ll 1)$ one can approximate Eq.~(1) as

\begin{equation}
T^R(\varepsilon)\simeq \frac{T_0}{R_0}\left(\frac{\varepsilon -\varepsilon_F}{D_0}\right)
^{2\gamma} .
\end{equation}
The physical meaning of the high-energy cutoff $D_0$ in Eqs.~(1) and (3) 
is the 
maximum energy transfered in a scattering event. For a strictly 1D wire with
a local electron-electron interaction (Luttinger liquid) a natural cutoff
is the bandwidth $ \varepsilon_F $. If the interaction potential is
characterized by a
finite length scale $l_s $ (a screening length) one has instead $\; 
D\sim \hbar v_F/l_s\;$. In a thin ($l_s { }_\sim^> W$) nanowire screening
effects could be suppressed. In this case the Coulomb potential
averaged over quantized transverse modes takes the form
$\;\; V_c(x)\simeq e^2/\sqrt{x^2+W^2}$
and the relevant high-energy cutoff is reduced to
$\; D_0\sim \hbar v_F/W\;$. 
The physics at scales $ x\ll W$ can not
be described by a long wavelength approximation and the Luttinger liquid-like
approach is inappropriate.

The reasoning outlined above leads to an expression for the 
nonlinear tunneling 
current which takes the following form ($eV \ll D_0$)

\begin{equation}
I_{NL} \simeq \frac{ev_F}{W}\frac{T_0}{R_0}
\left(\frac{eV}{D_0}\right)^{2\gamma+1} .
\end{equation}

We will model a nanocontact
formed in conductance quantization experiments as an atomic-size
constriction (characterized by a set of transmission coefficients
for quantized modes) attached to a long ($L\gg W$) nanowire of
width $W$ (see Fig.~1). In an adiabatic model of a nanoconstriction 
\cite{Shekhter} it is reasonable to
assume that $T_0$ is
{\em exponentially} close to zero or unity for all channels but one
(or possibly a few).
It should be stressed that in the range $\gamma \sim 1$ Eq.~(4) holds
even for $T_0$ close to unity due to a strong
renormalization of the transmission amplitude. The crossover value for the 
ratio of unrenormalized coefficients
$ T_0/R_0 \;$ that separates ``under-the-barrier" from ``over-the-barrier"
transport is $\; (T_0/R_0)_c \simeq (L/W)^{2\gamma}$ which is much bigger 
than unity 
if $\gamma \sim 1$. 
%
%In what follows we will refer to the electron transport
%through the constriction as being due to mesoscopic tunneling [so that 
%$T^R(\delta\varepsilon \ll D_0)\ll 1\;]\;$
%if $ \;T_0/R_0 \ll (T_0/R_0)_c\;$. 
%
If $T_0$ is close to one we should use the exact expression (1) rather than 
Eq.~(3) when describing high voltage behavior.
%
%for voltages close to and above the saturation 
%value $eV_s\sim D_0$.

\begin{figure}
\centerline{\psfig{figure=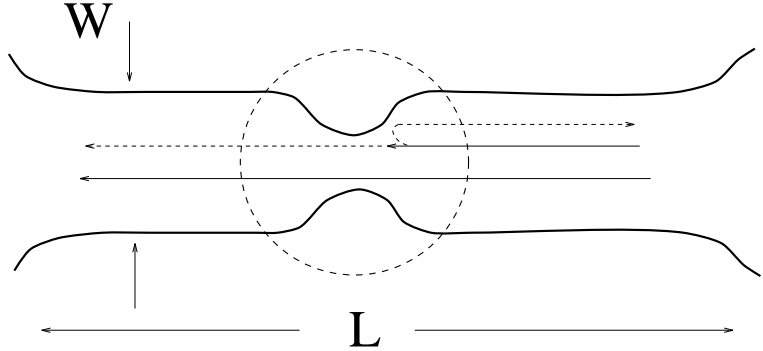,width=10cm}}
\vspace*{3mm}
\caption{\protect{\label{fig:geometry}}
Model of a nanowire contact. An atomic size constriction (within dashed
circle) characterized by a set of transmission coefficients for 
quantized modes is attached to a long nanowire of of length $L$ and
width $W$. Most modes are either fully transmitted (lower arrow) or
completely reflected (not shown) by the constriction. One, or possibly a few
modes are partially reflected (upper arrow).
}
\end{figure}

Since the nonlinear contribution to the currents measured
in Ref.~\cite{Garcia} was fitted to a power law, $V^q$, with an
exponent $q$ statistically peaked at $ q\simeq 3 $, the specific
value of interaction strength $\gamma = 1$ appears to be
of most interest to us. In this special case the nonlinear
conductance calculated from Eqs.~(1) and (2) takes a simple
analytic form
\begin{equation}
G(V) \equiv \frac{I(V)}{V} = \frac{e^2}{h}
\left[1-\frac{\arctan(x)}{x}\right],\; 
x=\frac{eV}{D_0}\sqrt{\frac{T_0}{R_0}}
\label{G}
\end{equation}
Therefore,
if the ``Coulomb blockade" of the quenched tunneling mode is lifted, 
the nonlinear voltage dependence of the current goes over to the linear 
dependence
expected for a perfectly transmitting mode.
As for the modes which are propagating, ``over-the-barrier" reflection
only results in a small additive renormalization of the linear
current. 
%In a context where we want to explain the nonlinear $I-V$ 
%characteristics of nanowires it is therefore only the ``tunneling modes" %which 
%are important for us.

Is it reasonable to trust the power law $I-V$ result even in cases when the
coupling strength is not small? We think so; actually this is a quite 
general result for
Luttinger liquid-like theories (see e.g. the 
reviews \cite{rev}). The
analogous expression to Eq.~(4) can be derived in the tunnel Hamiltonian
model for the whole range of interaction strengths. However, in this
approach one has to
assume a small unrenormalized transmission coefficient($T_0\ll 1$) to
get analytic results \cite{KF1,FN}. For our purpose it was important to
describe scattering of interacting electrons by a barrier of arbitrary height,
therefore we have used the approach elaborated in 
Ref.~\cite{Matveev_Glazman_Yue}.

For a finite wire of length $L$ renormalization of
the transmission amplitude should be stopped at energies
$\delta\varepsilon\equiv|\varepsilon - \varepsilon_F|
\leq\Delta = 2\pi\hbar v_F/L \;$. In the weak coupling limit the
low-energy cutoff $\Delta$ corresponds to the minimum-energy
charge excitations in a finite wire (longitudinal energy discretization).
In the case of strong interaction the cutoff depends on the
plasmon velocity $s$ and is
$\;\Delta(s/v_F)^2\sim e^2/L\;$ (see Section 3).
 When describing the  current in a finite wire Eqs.~(4) and (5) are
valid for voltages
%\begin{equation}
$\; V > V_c\simeq \Delta/e \;$ .
%\end{equation}
At lower voltages the transmission coefficient is energy independent
$ T^R(\delta\varepsilon\sim\Delta) = (T_0/R_0)(W/L)^{2\gamma}\;$.
The current produced by the quenched mode at low voltages, $V < V_c$,
is linear and small even if the bare transmission amplitude is close
to one,

\begin{equation}
I = \frac{e^2}{h}\frac{T_0}{R_0}\left(\frac{W}{L}\right)^{2\gamma} V,
\;\;\;\;\; V < V_c \; .
\end{equation}

\subsection{Comparison between theory and experiment}
Employing the above simple formulae for an explanation of the nonlinear
conductance of metallic nanowires we first note that the height of the
measured conductance
steps \cite{Garcia} deviate somewhat from multiples of $e^2/h$ 
(i.e. from perfect quantization). By associating the difference $\delta G$
with the small current carried by Coulomb blockaded mode at low voltages we  
conclude from Eq.~(6) that

\begin{equation}
\delta G/(e^2/h)\sim (T_0/R_0)(W/L)^{2\gamma} \, .
\end{equation}
As discussed above the suppression of the current in this quenched mode is 
lifted by a sufficiently large applied voltage. This gives rise to a 
nonlinear contribution 
to the $I-V$ curves when $V \sim V_c$. It follows that the nonlinearity
is due to
a suppression of Luttinger liquid-like effects at high voltages.
Since the nonlinear contribution to the current arises from the
 mode most sensitive to backscattering it does not depend on the number of conduction channels
(see below). The universal character of the nonlinear contribution is
due to the ``long distance" origin of barrier renormalization;
the relevant energy scale is either given by the total length of the nanowire
$L$ (if $V < V_C$ ) or by the ``field" length $l_E = \hbar v_F/eV \gg
W$.

Simple numerical estimations show that the above formulae allow us
to explain the $I-V$ characteristica measured in Ref.~\cite{Garcia}.
First of all the crossover voltage observed in the experiment
($V_c\simeq 0.1$~V) can be readily obtained from Eq.~(5) for
metallic nanowires a few nanometers long.  Since the numerical coefficient
in the theoretical estimation of $ V_c $ is uncertain it is difficult
to be more precise in the quantitative predictions. However, it is 
worthwhile
to stress that the voltage in question is determined by the
total length of the nanowire and not by the length of its narrowest
part which provides the conductance quantization. Therefore the
crossover to the nonlinear regime in this picture does not depend on
uncontrollable factors like precisely how the constriction is formed.

From the experiments of Ref.~\cite{Garcia} we conclude that the exponent 
$(2\gamma+1$) in Eq.~(4)
should be approximately equal to 3,  
hence $\gamma
%\simeq e^2/2\pi\hbar v_F 
\simeq 1$. For Coulomb interaction $V_{ee}(0)\gg V_{ee}(2k_F)$ and
one can estimate $\;\gamma\simeq (e^2/\pi\hbar v_F)\ln(L/W)\;$
which for reasonable values of $L/W$ and $v_F$ typical for
metals is in amazingly good agreement with the value extracted
from experiment. This numerical coincidence could be a strong
argument in favour of our ideas if both theoretical predictions
and the chosen ``experimental" value of exponent ($q=3$) in
power-law $I-V$ curves were {\em quantitatively} reliable.
Unfortunately there is no physical reason to believe that Eq.~(1)
is quantitatively correct in the range $\gamma\sim 1$. As for
experiment \cite{Garcia}, the distribution of the exponent $q$
for different measurements, although peaked at $q=3$,
spreads out from 2 to 4. So instead of playing with numbers
it would be more reasonable to look for {\em qualitative}
features which can be easily either confirmed or ruled out by
experiments. 

The qualitative predictions which emerge from our picture
are straightforward to extract from the theory. First, the larger the 
fluctuations of the conductance around its quantized value the bigger
the nonlinear current. One can find some experimental confirmation
of this prediction by examining Fig.~11 of Ref.~\cite{Garcia}. (The
fluctuations of conductance are obviously increasing as the
number of conduction channels [modes] increases. So one can expect that 
the absolute value of the nonlinear current will increase as well. Such a 
tendency can indeed be observed in Figs.~11(a-d) of the cited paper). Next, 
it is evident that the
contribution of the tunneling mode to the current when the Coulomb blockade 
is lifted by
 high voltages could at most be comparable to that of a freely propagating
mode (for which $ R_0 \ll T_0 $). Therefore the {\em total} current of 
an $ n $-channel constriction
at saturation voltage (which is determined by a short distance length scale) should approach the {\em linear} current
of an ($ n+1 $) channel nanocontact.

Plots of the nonlinear conductance $G(V)$ calculated from Eq.~(5) 
are shown in Fig.~2. Even for a very nearly ballistic mode
with $T_0=0.9$  one can see that there is a significant (although not complete)
restoration of transmittivity for voltages in a range relevant for the 
experiment. However for the quenched modes with small backscattering
complete restoration is achieved at high voltages
$V\sim V_s\simeq\sqrt{T_0/R_0}D_0\gg D_0$,
where the Luttinger liquid approach does not hold. In the theoretically
reliable region of voltages  $eV < D_0$ the current carried by
each individual (quenched) mode is a few times smaller than for a
perfectly transmitting channel. In the experiments
\cite{Garcia}, on the other hand, 
the conductance is increased by approximately one
quantum unit already in the low voltage region. In our picture the
observed conductance enhancement could be explained if more
than one suppressed mode contribute to the nonlinear current.
%This prediction is also supported
%(at least qualitatively) by the experiment (see Fig.~10 of %Ref.\cite{Garcia})
%if one supposes that the largest voltages used in \cite{Garcia}
%already correspond to the saturation value.

\begin{figure}
\centerline{\psfig{figure=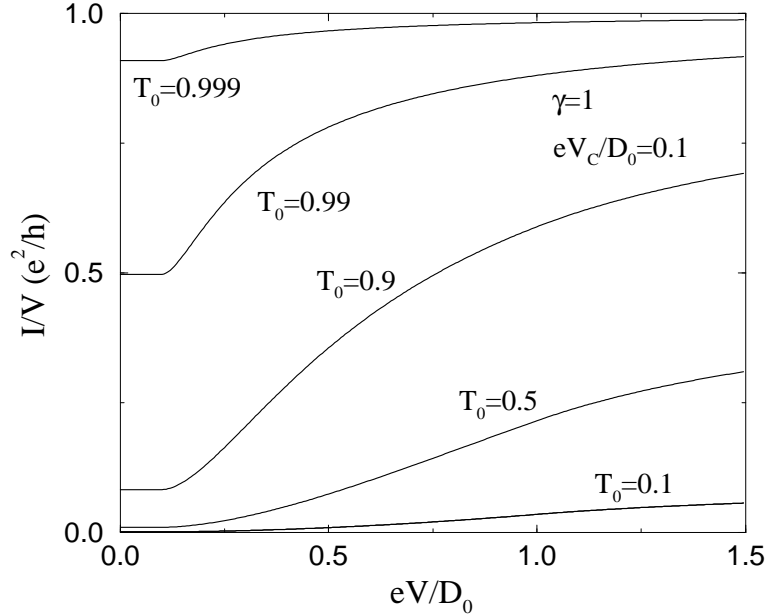,width=10cm}}
\vspace*{3mm}
\caption{\protect{\label{fig:GofV}}
Conductance $G(V)\equiv I/V$ calculated from the Tomonaga-Luttinger 
liquid model, Eqs.~(5) and (6). It represents the contribution of
a conductance channel in which electrons are
partly reflected ($T_0<1$) from a potential barrier
renormalized by electron-electron interactions, the strength of which is
characterized by a parameter $\gamma$. The parameter $D_0=\hbar v_F/W$
is a high energy cutoff associated with the time it takes for a charge 
excitation to spread across the width $W$ of the wire, while 
$V_c\sim \hbar v_F/L$ is a low energy cutoff related to the finite
wire length $L$.
}
\end{figure}

\subsection{Spin degrees of freedom and multichannel effects}
Although the above formulae pertain to the case of spinless fermions,
the inclusion of spin does not change the situation radically. As 
was shown in Ref. \cite{Matveev_Glazman_Yue}, adding the spin degree of 
freedom only results
in an obvious additional factor of 2 in the conductance for the case of a
smooth interaction potential, $ V_{ee}(2k_F)\ll V_{ee}(0) $. In a 
general case the transmission probability in Eq.~(1) should be replaced by
\cite{Matveev_Glazman_Yue}

\begin{equation}
T_0\left(\frac{\varepsilon}{D_0}\right)
^{2\gamma} \longrightarrow T_0 \left( 1 + 2\gamma_b\ln\frac{D_0}
{|\varepsilon-\varepsilon_F|}\right)^{3/2}\left|
\frac{\varepsilon-\varepsilon_F}{D_0}\right|^
{2\gamma_f -\gamma_b}
\end{equation}
It is evident that the additional logarithmic energy dependence of the
transmission probability can not change our estimations in an essential way.
%\footnote{Explain $\gamma_f$ and $\gamma_b$ in Eq. 9}

A less evident point in connection with 
the above purely 1D considerations is how they are
affected by multichannel effects. 
In reality a metallic nanowire may only be able to
support a few conductance channels at its most narrow part where the width
is of the order of a few Fermi wavelengths. 
The wire gets much broader, however, away from this region and many 
conduction channels are supported. To discuss this complication we will first
argue that even though the potential barriers that define the wire are 
due to atoms the adiabatic approximation applies.

In principle a rearrangement of the atomic structure as the the wire 
is extended could lead to a step-like
behavior of the conductance, but there is no reason to expect such
conductance steps to be quantized in units of $e^2/h$. 
Experiments, on the other hand, have
definitely demonstrated  such a conductance 
quantization (statistically) of metallic 
nanowires (see e.g. the review \cite{Jan}). Moreover, the
conductance staircase measured by the STM technique at room temperature 
\cite{Garcia} looks very similar to the one found in 2D electron gas systems. 
In particular, the
accuracy of the quantization is quite high (within 10\% for the first step) and
all plateaus were observed (up to $n = 6-8$) with no indication
of suppressed conductance steps at $n=2,4,7$ as predicted
%\footnote{What does this mean?}
for a cylindrically symmetric ballistic
microcontact. This means that for the selected measurements, and when the
conductance is quantized, the motion of electrons through the
constriction can be considered adiabatic; so a simple ``2D"-model
(with no degenerate transverse modes) of an adiabatic
contact \cite{Shekhter} may be quite adequate for our purposes.
%\footnote{I don't 
%understand this distinction between 3D and 2D contacts!?}

In the adiabatic approach to electron transport through a quantum
contact the transverse (fast) and longitudinal (slow) motion of 
electrons through the contact are decoupled when the Schr{\"o}dinger
equation is solved. The kinetic energy
due to the transverse motion then appears as a potential barrier for the
longitudinal motion which adds to the smooth potential from the constriction
itself. The smooth effective potential is different for different modes
and has a maximum where the contact diameter has its minimum. 
Since the energy level spacing imposed by transverse momentum (mode)
quantization
is of the order of the Fermi energy, one can use the following simple
classification of modes: those with energies above the 
barrier are transmitting modes, those below the barrier reflected
modes. As we have seen already this classification should be
changed for interacting electrons, where the renormalized potential
 rather than the 
bare potential is relevant. One can infer from Eq.~(1) that
all modes for which $ \; T_0/R_0 \ll (L/W)^{2\gamma}\gg 1 \; $ are 
quenched by the strong enhancement of backscattering. 
It is known
that for smooth potentials over-the-barrier reflection  decreases
exponentially with increasing energy of the incident particle.
Hence it is quite natural to assume that only the last mode (the one with
energy closest to the potential barrier maximum) will change its character by renormalization effects and be
converted from a transmitting to a reflected mode. All other
modes keep their unrenormalized character; they are either
transmitting modes (over-the-barrier) or totally reflected modes 
(under-the-barrier).

Let us now ask the question how a transmitting mode with $T_0=1$
is affected by inter-mode (inter-channel) {\em forward} scattering in 
a multichannel wire. The answer is that in an effectively infinite 
 multichannel Luttinger-liquid wire the Coulomb
interaction smears out the quantized conductance plateaus 
so that in the strong interaction limit the 
conductance is described by almost smooth function of the wire width (or of the Fermi energy)
 \cite{Matveev}. This effect is entirely
due to the properties of plasmon excitations at long distances. 
%\footnote{Over long distances?? long wavelength plasmons??}. 
For a wire 
of {\em finite} length connected to large reservoirs the dc current will
be determined by the properties of plasmons in the reservoirs \cite
{Matveev,Maslov_Stone}. So in experiments, where the reservoirs
are massive metals, the conductance of a perfect multichannel wire of 
interacting
electrons will be quantized exactly as if the electrons in the wire were 
noninteracting.
%The only effect of interaction is to further suppress the 
%small correction due to  over-ther-barrier reflection. It is interesting
%to speculate that the relatively high accuracy of conductance %quantization 
%observed in metallic nanowires could be explained by this very effect.

Now we turn to another question: Could the Coulomb-blocked mode, 
which is reflected by the renormalized 
barrier, affect the propagation of the transmitting modes? In the
case of weak interaction the answer is that it definitely cannot. First of
all the presumed smooth bare potential can not induce intermode
scattering. Therefore each mode with $\; R^j_0 \neq 0\;$ ($ j$
is the mode index) sets up its own Friedel oscillations when reflected
by the potential. The corresponding induced barriers are characterized 
by different oscillation periods $\; \lambda(\varepsilon =\varepsilon^j_F)\;$, 
where $\; \varepsilon^j_F\;$ is the ``Fermi energy" in the $j$:th channel.
(The channel dependence comes from the fact that $\varepsilon^j_F$ is measured
from the quantized transverse energy corresponding to the $j$:th mode; it
is the bandwidth of the $j$:th mode).
Interaction enhanced backscattering arises only for electrons in the same
mode or subband [since the corrections to bare transmission amplitude is
proportional to $R_0^{(j)}$].
 Particles with energies in the vicinity of $\;\varepsilon^j_F\;$
are phase locked to the  Friedel oscillations and thus strongly
backscattered. The influence of this effect on the transmittivity of
quantized modes have already been studied above.
Perfectly transmitting modes ($R_0^{(j)}=0$) can be scattered by the
potential due to Friedel oscillations of the 
charge density in the quenched channel. However, this scattering
could lead at most to nonsingular (length independent) contributions.
In the lowest order of perturbation theory the corresponding
induced backscattering amplitude $\;|r^{(j)}|=\pi\gamma |r_0|\;(\;r_0$
is the bare backscattering amplitude of quenched mode) is small
if $|r_0|\ll 1$.

  To put it differently, transmitting
modes will propagate freely even in an interacting electron system
irrespective of possibly strong backscattering in the other modes.
Is it possible to extend this claim to the strongly interacting case? 
To clarify this
point we switch now to the case of strong coupling, where the picture
of Luttinger liquid transport through a constriction will look
slightly different.

\section{The strong coupling regime}
In the strong coupling regime the suppression of tunneling for repulsively
interacting electrons can be easily understood in terms of a
Wigner crystal (or charge density wave) pinned by impurities
\cite{GRS}. The quantum depinning of a 
Wigner crystal is accompanied by propagation
of plasmon excitations along the wire. The gapless sound-like
spectrum of 1D plasmons results in a logarithmically divergent
contribution of plasmons to the tunnel action. The infrared cutoff
(at small momenta) appears at scales determined by temperature
or voltage. However, in the case of a {\em finite} wire (either
isolated or connected to 3D leads) one gets the additional natural energy scale
$\Delta_s = \hbar s/L$, where $s$ is the plasmon velocity. For an isolated
wire (or a ring, see e.g. \cite{Krive}) this scale is provided by
the discrete spectrum of energy levels; for a wire connected to 3D
leads a cutoff appears due to the fact that the excitation spectrum
of 3D plasmons has a gap and does not give rise to infrared divergencies.

As in the case of weak coupling we will associate the nonlinear contribution
to the current with the interaction-quenched mode.
If the interaction is not weak one can treat the
Luttinger liquid transport through a barrier analytically only in the
limit of a strong barrier. So our consideration could be
applied strictly speaking to the tunneling regime of conductivity.
We start with a purely 1D picture (a single mode channel). Now the
current at temperatures effectively equal to zero (for metals this
temperature interval includes room temperature) is given by

\begin{equation}
I(V) = \frac{e}{h}\int_0^{eV} d\varepsilon D(\varepsilon),
\end{equation}
where

\begin{equation}
D(\varepsilon) = \left(\frac{\varepsilon}{V_p}\right)^{\frac{2}{\alpha}},\;\;\;
\alpha = \frac{v_F}{s} \ll 1 \; .
\end{equation}
is the tunneling probability associated with the excitation of long
wavelength plasmons \cite{GRS} (notice that in this reference the
exponent in the formula for the tunneling probability contains an incorrect
numerical factor). In Eq.~(10) $V_p$ is the pinning potential which
for strong repulsion ($\alpha\ll 1$) can be put equal to the height
of the potential barrier \cite{GRS}.

At high voltages $eV >\Delta_s/\alpha $ (however still much lower than
the height of the barrier) the current 
is  nonlinear

\begin{equation}
I_{NL}(V) \simeq \frac{e}{h}\alpha V_p
\left(\frac{eV}{V_p}\right)^{\frac{2}{\alpha}} \; .
\end{equation}
Notice that we omitted terms of order unity in the exponents of Eqs.~(10)
and (11) since the accuracy of calculations in the strong coupling limit 
($\alpha\ll 1$) is not so good as to make it meaningful to keep them.

At small voltages the tunnel action should be cut off at the energy
scale $\Delta_s$ and the tunnel current becomes linear

\begin{equation}
I = \frac{e^2}{h}\left(\frac{\Delta_s}
{V_p}\right)^{\frac{2}{\alpha}}V,\;\;
eV \ll \Delta_s/\alpha \; .
\end{equation}

To get the kind of nonlinear $I-V$ characteristics measured in 
Ref.~\cite{Garcia} one should put $\alpha\simeq 2/3$ in Eq.~(12). 
Again, this value of the
correlation parameter is not sufficiently small to justify a
quantitative comparison between theory and experiment.

An equation for the nonlinear current [analogous to Eq.~(11)] valid in
the required range of interaction strength can be derived in the tunnel
Hamiltonian approach \cite{FN}. The corresponding formula for a {\em finite}
wire connected to 1D reservoirs of {\em noninteracting} electrons
(for ballistic transport an analogous model was considered in 
Ref.~\cite{Maslov_Stone}) takes the form

\begin{eqnarray}
I(V) & = & \frac{4e}{\hbar}\frac{|t|^2}{\Delta_s}
\int_0^{\infty}dx\cos\left\{\frac{eV}{\Delta_s}x - \frac{2}{\alpha}
\arctan\left(\frac{\Lambda}{\Delta_s}x\right) + \left(\alpha^{-1}-1\right)
\left[\pi +2si(x)\right]\right\} \nonumber \\ 
&\times & \left[1+\left(\frac{\Lambda}{\Delta_s}x\right)^2\right]
^{-\frac{1}{\alpha}} 
 \exp\left\{2\left(\alpha^{-1}-1\right)\left[C+\ln(x)-
ci(x)\right]\right\} ,
\label{14}
\end{eqnarray}
where $t$ is the tunneling matrix element which is assumed to be small,
$C\simeq 0.577\ldots$ is the Euler constant,
$\Lambda$ is the ultraviolet cutoff ($\Lambda\gg\Delta_s,\;
\Lambda\gg eV$), and the integral sine (cosine) function is  
defined as follows

\begin{equation}
\left(\begin{array}{c} si \\ ci \end{array}\right)(x) = -\int_x^{\infty}dt\frac{\left(\begin{array}{c} \sin \\ \cos \end
{array}\right)(t)}{t}
\label{15}
\end{equation}

For $\Delta_s\rightarrow 0$ (infinite wire) Eq.~(\ref{14})
yields the well-known result for Luttinger liquid transport
of a power-law $I-V$ dependence \cite{KF1,FN}

\begin{equation}
I_{NL} \simeq \frac{e}{h}\frac{|t|^2}{\Lambda}
\left(\frac{eV}{\Lambda}\right)^{2\alpha^{-1}
-1} \; .
\label{16}
\end{equation}
For a finite wire the integral in Eq.~(\ref{14}) has to be
evaluated numerically. The corresponding $I-V$ curves are
plotted in Fig.~3. From this figure one can see that
at low voltages the current grows {\em linearly with voltage} and is small. 
Beginning
from $ V\sim V_{c1}\simeq\Delta_{s}/e $ the tunnel current 
 increases rapidly, eventually reaching the asymptotic value Eq.~(\ref{16})
at $ V > V_{c2}\simeq \Delta_{s}/e\alpha $.

\begin{figure}
\centerline{\psfig{figure=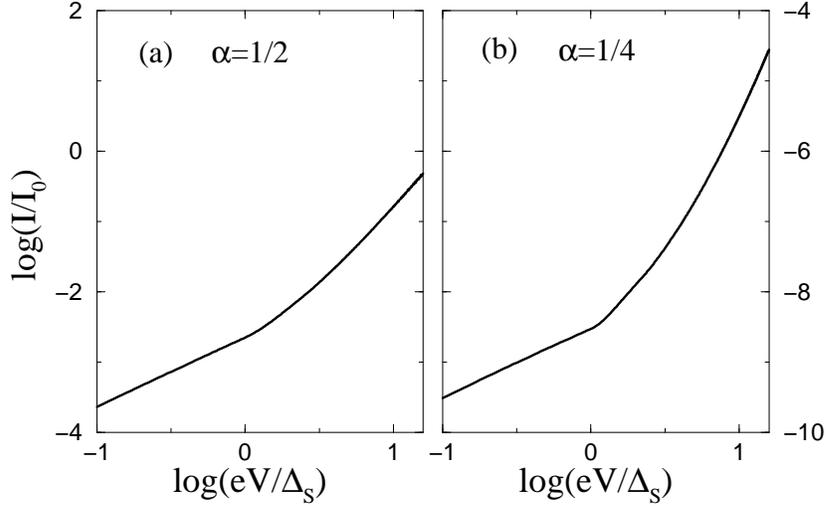,width=12cm}}
\vspace*{3mm}
\caption{\protect{\label{fig:logI_vs_logV}}
Log-log plots of the nonlinear tunneling current (Eq.~13) 
vs. gate voltage for a 
finite quantum wire connected to reservoirs of non-interacting electrons. 
Here $I_0 = (e/h)|t|^2/\Lambda$, where $t$ is the tunneling matrix element, 
$\Lambda \sim \hbar v_F/W$ is a high-energy cut-off related to the wire 
width $W$; $\Delta_s = \hbar s/L$ ($s$ is the plasmon velocity) is an energy
scale associated with the discrete energy spectrum and hence the finite
wire length $L$. In the calculation 
$\Lambda/\Delta_s=50$. For small voltages, $V<\Delta_s/e$,
 the current increases linearly with voltage. For higher voltages the
current depends on the interaction parameter $\alpha$ and the slope in
the figure approaches the infinite-wire length value $2/\alpha-1$ of Eq.~(15).
}
\end{figure}

As we have noticed already the above formulae describe the
tunnel current and strictly speaking they can not be used
for explanaining the nonlinear $I-V$ behavior measured
in the experiment \cite{Garcia} (where the observed
increase of conductance was of the order of the conductance
quantum). However, in order to analyze the phenomenon in terms relevant
for the strong coupling regime one may put $\; |t|\sim\Lambda\sim
\hbar v_F/W\;$ . For a comparison with experiment it is reasonable
to use the expression [analogous to Eq.~(\ref{16})],
which describes the nonlinear current of spin-1/2 electrons.
The corresponding formula can be readily obtained from Eq.~(\ref{16})
by the substitution $\; 2\alpha^{-1}\rightarrow
\alpha^{-1}_{\rho}+\alpha^{-1}_{\sigma}\;$, where
$\; \alpha_{\rho}\;(\alpha_{\sigma})\;$ is the Luttinger liquid correlation
parameter for the charge (spin) sector).
For an $SU(2)$ invariant spin
interaction (relevant here) $\;\alpha_{\sigma} = 1\;$. The correlation
parameter in the charge sector can be expressed in terms of the Fourier 
transform of the forward scattering interaction potential 
(see e.g. Ref.~\cite{Matveev})

\begin{equation}
\alpha^{-2}_{\rho} = 1 + \frac{2V_{ee}(0)}{\pi\hbar v_F} \; .
\end{equation}

For a cubic $I-V$ behavior $\;\alpha_{\rho} \simeq 1/3\;$
and the corresponding
value of the dimensionless interaction strength is $\;\gamma\simeq 2\;$,
which is larger by a factor 2 than the value extracted in the weak coupling
limit. All numerical estimations performed for the weak coupling 
limit can be equally well
reproduced using formulas corresponding to the strong coupling regime 
(obviously with the same uncertainties
in numerical coefficients of order one). As a result one
has to admit the evident failure to {\em quantitatively} describe
nonlinear transport in metallic nanowires using naively a Luttinger
liquid model. However it is suggestive that both limits yield
reasonable values for the interaction parameters (of order one, which is
typical for metals). 

The advantage of looking at our problem in the strong
coupling limit is that it allows us to consider the
multichannel case analytically. 
At first we assume that the current through the nanoconstriction
is due to tunneling. Then the problem is reduced to describing the propagation
of charged excitations (plasmons) in a multichannel wire.
For this case (here we again for simplicity will consider spinless fermions) 
it is known \cite{Matveev} that in the limit of a large 
number of channels $N_{\perp}\gg 1$ the plasmon velocity $s$ is given by

\begin{equation}
s = \sqrt{\sum_{j=1}^{N_{\perp}}n_j\frac{e^2}{m}} \; ,
\end{equation}
where $n_j$ is the electron density in the $j$-th channel. Hence it is
determined by the {\em total} density of electrons in a multichannel wire.
It is this very density (which approaches the bulk density of electrons in the 
vicinity of the electrodes) and not the smaller density of 
electrons in the few quantized modes 
at the center of the contact which
 will determine the actual value of the correlation
parameter $\alpha$ in Eq.~(11). Therefore, the exponent in the power-law type
$I-V$-curves should be independent of the number of conductions channels
as was claimed in Ref.~\cite{Garcia}.

A situation where the number of channels varies along the wire can  be modelled
 by assuming the plasmon velocity to be a function of
position $ s\rightarrow s(x) = s_{0}f(x)$. Straightforward
 calculations analogous
to those performed in Ref.~\cite{GRS} (see also \cite{Krive}) yield

\begin{equation}
D(\varepsilon) = \exp\left(-\frac{2}{\alpha_0}\int_{\varepsilon}^{V_p}
dx\frac{f(x)}{x}\right) \; .
\end{equation}
According to Eqs.~(9) and (18) the nonlinear conductance of a general 
multichannel wire will depend on the actual shape of the wire. However,
 if the function
$s(x)$ is smooth, the tunneling probability still has a local
energy dependence,

\begin{equation}
D(\varepsilon) = \left(\frac{\varepsilon}{V_p}\right)^{\frac{2}{\alpha(\varepsilon)}},
\;\; \alpha(\varepsilon) = v_F/s(\varepsilon) \; .
\end{equation}
Therefore to get in this approach a sample-independent exponent describing the
nonlinear conductance observed in Ref.~\cite{Garcia} one has to
assume that the nanowire has an almost
uniform shape far from the constriction (this is exactly the
model for a nanocontact used here, see Fig.~1).

The last point we would like to discuss is how the mode
quenched by interaction effects influences the transmitting modes.
In the strong coupling regime the assumption of perfect transmission
means that the corresponding Wigner crystals can slide freely (without
distortions) along the channels. If the interaction potential is smooth
and one can neglect effects of backscattering [which are proportional
to the exponentially small interaction constant $V(2k_F)$] the degrees
of freedom associated with perfectly transmitting modes decouple
from the rest. Their dynamics is linear and can be
treated exactly \cite{Maslov_Stone} for an adiabatic constriction.
One may conclude that as long as the backward scattering part of
the interaction potential is irrelevant, perfect conductance quantization
can not be spoilt by quenching of reflected modes.

\section{Conclusions}
Summing up all facts in favour of the above picture, we conclude that the
Luttinger liquid-like effects could be a reasonable 
explanation of the nonlinear
conductance measured in metallic nanowires. If so the experiment by
Costa-Kr{\"a}mer {\em et al.} 
\cite{Garcia} can be regarded as one of only a few highly important experiments
(see also \cite{Webb,Chang}) where the fundamental
properties (Kane-Fisher effect) of Luttinger liquid-like states of strongly
correlated electrons were observed.

\section*{Acknowledgements}
It is a pleasure to dedicate this work to Rolf Landauer
on the occasion of his 70:th birthday. We are grateful to Nicolaus Garc{\'i}a
and H{\aa}kan Olin for bringing the puzzle of the origin of the nonlinear
nanowire $I-V$ characteristics measured in Ref.~\cite{Garcia} to our attention 
and for valuable discussions.
This work was supported by the EU ESPRIT Project 
Nanowires, by the Swedish Royal Academy of Sciences (KVA) and by the Swedish 
Natural Science Research Council (NFR).

\label{lastpage}


\begin{thebibliography}{0}
\bibitem{channels}
Quantization of the transverse electron motion in the wire gives rise to 
modes which are separated in energy. In our terminology each occupied
mode gives rise to one conduction channel.
\end{thebibliography}
\end{document}